\documentclass[12pt]{article} 
\usepackage[sectionbib]{natbib}
\usepackage{array,epsfig,fancyheadings,rotating}

\usepackage[]{hyperref}  

\usepackage{sectsty, secdot}
\sectionfont{\fontsize{12}{14pt plus.8pt minus .6pt}\selectfont}
\renewcommand{\theequation}{\thesection\arabic{equation}}
\subsectionfont{\fontsize{12}{14pt plus.8pt minus .6pt}\selectfont}

\usepackage{amsmath}
\usepackage{amssymb}
\usepackage{amsfonts}
\usepackage{multirow}
\usepackage{amsthm}

\newtheorem{theorem}{Theorem}
\newtheorem{lemma}{Lemma}
\newtheorem{corollary}{Corollary}

\theoremstyle{definition}

\begin{document}


\renewcommand{\baselinestretch}{2}

\markright{ \hbox{\footnotesize\rm 
}\hfill\\[-13pt]
\hbox{\footnotesize\rm
}\hfill }

\markboth{\hfill{\footnotesize\rm Xietao Zhou and Steven Gilmour} \hfill}
{\hfill {\footnotesize\rm FILL IN A SHORT RUNNING TITLE} \hfill}

\renewcommand{\thefootnote}{}

$\ $\par


\fontsize{12}{14pt plus.8pt minus .6pt}\selectfont \vspace{0.8pc}
\centerline{\large\bf $Q_B$-OPTIMAL TWO-LEVEL DESIGNS}
\vspace{2pt} 
\centerline{\large\bf FOR THE BASELINE PARAMETERIZATION}
\vspace{.4cm} 
\centerline{Xietao Zhou and Steven Gilmour} 
\vspace{.4cm} 
\centerline{\it King's College London}
 \vspace{.55cm} \fontsize{9}{11.5pt plus.8pt minus.6pt}\selectfont


\begin{quotation}
\noindent {\it Abstract:}
We have established the association matrix that expresses the estimator of effects under baseline parameterization, which has been considered in some recent literature, in an equivalent form as a linear combination of estimators of effects under the traditional centered parameterization. This allows the generalization of the $Q_B$ criterion which evaluates designs under model uncertainty in the traditional centered parameterization to be applicable to the baseline parameterization. Some optimal designs under the baseline parameterization seen in the previous literature are evaluated and it has been shown that at a given prior probability of a main effect being  in the best model, the design converges to $Q_B$ optimal as the probability of an interaction being in the best model converges to 0 from above. The  $Q_B$ optimal designs for two setups of factors and run sizes at various priors are found by an extended coordinate exchange algorithm and the evaluation of their performances are discussed. Comparisons have been made to those optimal designs restricted to level balance and orthogonality conditions.

\vspace{9pt}
\par
\end{quotation}\par

\def\thefigure{\arabic{figure}}
\def\thetable{\arabic{table}}

\renewcommand{\theequation}{\thesection.\arabic{equation}}

\fontsize{12}{14pt plus.8pt minus .6pt}\selectfont

\section{Introduction}
{
Factorial designs of various types are useful when we are studying the effects of several factors on one or more responses, and they have seen wide application in industrial research and many other areas. Traditional approaches assume that the model that will be fitted to the data is fixed and some classic optimality criteria have been applied to evaluate the designs under this model. Early extensions have been made so that a few alternative models could be considered as in \cite{Nachtsheim}. The $Q_B$ criterion has then been proposed to offer the capacity of considering hundreds of alternative models that could potentially be useful for data from a multifactor design in \cite{Factorscreening}. This method avoids the risk of going to the extreme belief that the model proposed is the best and allows experimenter’s beliefs about the possible models to be reflected in the design selection process by assigning different prior probabilities to each possible model. 

Follow up work which generalized the $Q_B$ criterion to different numbers of levels of factors, various kinds of maximal model and qualitative or quantitative factors can be found in \citet{Gilmour2010} and optimal $Q_B$ designs under the first order maximal model obtained by conference matrices have been given in \cite{saturated}. The conditions for a level balanced design, i.e.\ each column has the same number of ‘+1’ and ‘-1’ in the design, to be $Q_B$ optimal has been discussed in \cite{saturated} as well.

Recently an alternative parameterization of models for factorial experiments, called the baseline parameterization, has received considerable attention in the literature. It has been argued in \cite{MukerjeeRahul2012} that this parameterization arises quite naturally in many practical situations. That paper proposed the minimum K-aberration criterion to search for optimal designs; it sequentially minimizes the bias from the factorial effects of order $i \in \{2,\ldots,m\}$ when estimating the main effects, where $m$ is the number of factors in a design and we will define the order of a factorial effect in Section~\ref{sec:link}. 

It has been proved in \citet{Gilmour2010} that the centered parameterization version of the minimum generalized aberration criterion, $G_2$ aberration introduced in \cite{deng1999minimum}, is a limiting form of the $Q_B$ criterion. The present work aims to generalize this idea to show that a version of the $Q_B$ criterion applicable to the baseline parameterization, could not only include the case when the minimum K-aberration design in \cite{MukerjeeRahul2012} is indeed optimal, but also extend it to other potential optimal designs under model uncertainty not considered by \cite{MukerjeeRahul2012}, thus offering the experimenter additional flexibility.

The rest of the paper is organized as follows. In Section~\ref{sec:link} we derive the explicit relationship between the estimators of effects under baseline and centered parameterizations. The $Q_B$ criterion that is applicable to the baseline parameterization is then proposed in Section~\ref{sec:expression}. Using the $Q_B$ expression obtained we evaluate the existing minimum K-aberration designs from \cite{MukerjeeRahul2012} in Section~\ref{sec:evaluation}. The  $Q_B$ optimal designs for two setups of numbers of factors and run sizes for various priors found by an extended coordinate exchange algorithm, together with an assessment of their performance and comparisons with optimal designs restricted to level-balanced and orthogonal conditions, will be given in Section~\ref {sec:optimal}. The conclusions and discussion are given in Section~\ref{sec:conclusion}.
}
\section{The link between estimators under baseline and centered parameterizations}
\label{sec:link}

In this section, we derive the explicit formulae linking the baseline parameterization, which has been considered in some recent literature, e.g.\ in \cite{MukerjeeRahul2012} and \cite{ZhangRunchu} and the traditional centered parameterization. The baseline parameterization encodes the factorial effects in the design matrix with ‘1’ and ‘0’, corresponding to the high level ‘+1’ and the low level ‘-1’ respectively of the factor under the centered parameterization (which is also called the orthogonal parameterization in some literature such as \cite{Cheng-YuSunandBoxinTang}). We arrange the factorial effects in Yates' order which can be defined as follows: $a_0$=\textbf{1} denotes the list of factorial effects where only the intercept is considered; $a_k$=\{$a_{k-1}$, $F_k\odot a_{k-1}$\} for all $k \in \{1, \ldots, m\}$, where $\odot$ denotes the elementwise product of factorial effect of factor $k$ denoted by $F_k$, and all the other effects contained in $a_{k-1}$.
Let $\hat{\theta}$ and  $\hat{\beta}$ denote the vectors of estimators of the factorial effects in Yates' order under the baseline and centered parameterizations respectively. Then we can obtain an expression $\hat{\theta}$=$A_m$ $\hat{\beta}$ where $A_m$ represents the linkage matrix when $m$ factors are involved as given by the following theorem.

\begin{theorem}
\label{thm:theorem 1}
Under an $m$-factor full factorial  unreplicated design, the general form of the matrix $A_m$ is given by
$A_0=1$ and 
\[A_m= \begin{bmatrix} A_{m-1}&-A_{m-1}\\ \mathbf{0}&2A_{m-1} \end{bmatrix} ,\] 
for all $m > 0$ and for factorial effects arranged in Yates' order.  
\end{theorem}

Before we give a proof of this theorem, we will first present and prove a lemma regarding the general structure of the design matrices $X_{b(m)}$ and $X_{c(m)}$, the design matrices for $m$ factors under baseline and centered parameterizations respectively.

We adopt a specific way of arranging the rows and columns in the design matrix $X$ such that the factorial effects given by the columns of the design matrix appear in the Yates order and the level combinations of these factorial effects given by the rows of the design matrix are determined through the reverse of Yates' order. For the two factor model as an example, its design matrix $X_{b(2)}$ under the baseline parameterization is
\[
\begin{bmatrix} 1&1&1&1\\1&0&1&0\\1&1&0&0\\1&0&0&0 \end{bmatrix}. 
\]
The first row corresponds to the level combination of effects  where both factors are set to the high level, the second row corresponds to setting only $F_2$ to the high level, the third row corresponds to setting only $F_1$ to the high level and the final row corresponds to the case where both factors are set at the lower level. This gives the right order of the rows of level combinations.

Under this set-up, we are able to prove the following lemma regarding the structure of the design matrices in the $m$-factor case for $X_{b(m)}$ and $X_{c(m)}$ respectively.

\begin{lemma}
\label{thm:lemma 1}
The general form of the design matrices $X_{b(m)}$ and $X_{c(m)}$ can be given by \[X_{b(m)} =
 \begin{bmatrix} X_{b(m-1)}&X_{b(m-1)}\\X_{b(m-1)}& \mathbf{0}\end{bmatrix} ,\] and \[X_{c(m)}= \begin{bmatrix} X_{c(m-1)}&X_{c(m-1)}\\X_{c(m-1)}& -X_{c(m-1)}\end{bmatrix} ,\] given that the set up of the design matrix $X$  is as above.\par
\end{lemma}

\textit{Proof}:
Consider the design matrix $X_{b(m)}$. The first $2^{m-1}$ factorial effects are ordered in the sequence $a_{m-1}$ and the rest of the factorial effects are given by $F_m\odot a_{m-1}$. Since $F_m$ is set to the lower level 0 in the bottom half of the rows in the design matrix according to our set-up, the bottom right block matrix gives a $\textbf{0}$ matrix. On the other hand, $F_m$ is set to the high level 1 in the top half of the rows in the design matrix, so the top right block matrix gives $X_{b(m-1)}$. Also, the set-up of factor $F_m$ does not affect the full design matrix of the previous $2^{m-1}$ factorial effects, which means that the top left block matrix reads $X_{b(m-1)}$, and thus our result follows. A similar argument works for the design matrix $X_{c(m)}$, a slight difference being that the lower level of factor $F_m$ is -1 instead of 0 in the bottom rows of the design matrix $X_{c(m)}$, so that the bottom right block matrix will then be $-X_{c(m-1)}$.\par

We may now proceed to the proof of Theorem~\ref{thm:theorem 1}.\par

\textit{Proof}:
In an unreplicated full factorial design, we can rewrite the expressions of the least squares estimators of ${\theta}$ and ${\beta}$ to get a new expression of the form $\hat{\theta}=T\hat{\beta}$ where $T$=$X_{b(m)}^{-1}X_{c(m)}$ (this is because in an unreplicated full factorial design, the design matrix $X$ is a square matrix and every row and every column is linearly independent from other rows and columns, therefore the design matrix is invertible). If we know the general form of $X_{b(m)}^{-1}$, Theorem~\ref{thm:theorem 1} will then be proved. A standard result on 2$\times$2 block matrices' inverses has been applied to 2$\times$2 triangular block matrices using results summarized by \cite{lu2002inverses}.

Consider the triangular 2$\times$ 2 block matrix 
\[
\begin{bmatrix} A&B\\C&\textbf{0} \end{bmatrix},
\]
When the block matrices A, B  and C are all square matrices of the same dimension and invertible,  the inverse of 
\[
\begin{bmatrix} A&B\\C&\textbf{0} \end{bmatrix}
\]
is given by
\[
\begin{bmatrix} \textbf{0}&C^{-1}\\B^{-1}&-B^{-1}AC^{-1} \end{bmatrix}
\]
as shown by \cite{lu2002inverses}.

In our case, A=$X_{b(m-1)}$, B=$X_{b(m-1)}$, C=$X_{b(m-1)}$ and therefore, as discussed above, 
\[
X_{b(m)}^{-1}= \begin{bmatrix} \textbf{0}&X_{b(m-1)}^{-1}\\X_{b(m-1)}^{-1}&-X_{b(m-1)}^{-1} \end{bmatrix}.
\]

We are now ready to plug in all we have obtained into our formula $A_m$=$X_{b(m)}^{-1}$$X_{c(m)}$ and by the standard result of 2$\times$2 block matrices' multiplication, Theorem~\ref{thm:theorem 1} is then proved as $A_{m-1}$=$X_{b(m-1)}^{-1}$$X_{c(m-1)}$.

We note that an analogous result to Theorem~\ref{thm:theorem 1} has been seen in \cite{Cheng-YuSunandBoxinTang}, but there are two differences between our framework and theirs. First, we have adopted a more natural mapping between the two parameterizations, using the lower level of the centered parameterization mapping to the lower level of the baseline parameterization and similarly for high levels. Furthermore, the philosophy between our frameworks are different. While in \cite{Cheng-YuSunandBoxinTang} the aim is to find designs using the baseline parameterization, we are trying to use our result to include everything under the centered parameterization, which has been studied more thoroughly, while preserving the potential usefulness of baseline parameterization.

For the purpose of this paper, we will primarily focus on factorial designs with unreplicated design points which can be taken from an unreplicated full factorial design. Still, it shall be interesting to know that the above can be generalized to replicated full factorial designs for potential further research as in the following corollary.

\begin{corollary}
\label{thm:corollary 1}
If all the treatments are replicated $r$ times, $A_m^{'}$ in the replicated case will be the same as $A_m$ in the single replicated case.
\end{corollary}

\textit{Proof}:
We know by Theorem~\ref{thm:theorem 1} that $\hat{\theta}$=$A_m$$\hat{\beta}$ for a specific choice of matrix $A_m$. We may rewrite the expression as $(X_b^TX_b)^{-1}X_b^T=A_m(X_c^TX_c)^{-1}X_c^T$ and we aim to find a $A_m^{'}$ such that a similar expression holds in the replicated case.

If all the treatments are replicated $r$ times, 
\[
X_b^{'}=  \begin{bmatrix} X_b\\X_b\\ \vdots\\X_b\end{bmatrix}
\]
and the corresponding $(X_b^T)'$=$\begin{bmatrix} X_b^T&X_b^T \cdots&X_b^T\end{bmatrix} $. In both cases, the sub-block matrices are replicated $r$ times. We can then calculate $((X_b^T)'X_b^{'})^{-1}$= $(X_b^TX_b)^{-1}/r$ and analogous results hold for the centered parameterisation. Thus in the end we can obtain 
\[
\begin{bmatrix}(X_b^TX_b)^{-1}X_b^T&(X_b^TX_b)^{-1}X_b^T\cdots&(X_b^TX_b)^{-1}X_b^T\end{bmatrix}
\]
\[
=
A_m^{'}\begin{bmatrix}(X_c^TX_c)^{-1}X_c^T&(X_c^TX_c)^{-1}X_c^T\cdots&(X_c^TX_c)^{-1}X_c^T\end{bmatrix},
\]
which implies that $A_m^{'}$=$A_m$.

We define a factorial effect of order $i$ as the effect that consist of $i$ factors. Under this definition, the intercept is the effect of order 0, main effects have order 1; two-factor interactions have order 2, etc. This gives the following corollary.

\begin{corollary}
\label{thm:corollary 2}
Under the previous set-up, if $\hat{\beta}$'s order i and higher order terms are 0, where $i=0, 1, 2, \cdots, m$, then $\hat{\theta}$'s terms of order i and higher are 0.
\end{corollary}

\textit{Proof}: By Theorem~\ref{thm:theorem 1}, for all $m$, expanding out matrix $A_m$ and writing a system of equations, it is always true that the $\hat{\theta}$'s of order $i$ can be written as a linear combination of $\hat{\beta}$'s of order $i$ and higher order terms up to order $m$. Therefore if $\hat{\beta}$'s order $i$ and higher order terms are 0, then $\hat{\theta}$'s order $i$ terms are 0. Moreover, the linear equation for $\hat{\theta}$'s order $i$ terms contains all the terms of $\hat{\beta}$ in the linear equation of $\hat{\theta}$'s order higher than $i$ terms, Corollary~\ref{thm:corollary 2} follows.

By Corollary~\ref{thm:corollary 2}, up to the interchange of columns and rows within the association matrix $A_m$, if we have a maximal model of up to second order effects, we can then obtain our association matrix $A_m^{''}$ by shrinking our original 
association matrix $A_m$ to a block matrix up to but not including all rows and columns of effects of order 3. This leads to the derivation of the $Q_B$ expression under the baseline parameterization setting in the next section.

\section{Expression for $Q_B$ under the baseline parameterization}
\label{sec:expression}
In \cite{Factorscreening} the $Q_B$ criterion function was defined as 
\begin{equation}\label{3.1}
Q_B=\sum_{s=1}^{n_0} \sum_{i=1}^v \sum_{j=0}^v r_{i j} M_s(i, j) \tilde{\operatorname{Pr}}\left(M_s\right).
\end{equation}
Here, $r_{i j}=\frac{a_{i j}{ }^2}{a_{i i}{ }^2 a_{j j}}$ where $a_{i j}$ are the elements of the information matrix for the maximal model under the centered parameterization, $n_0$ is the total number of eligible submodels nested in the maximal model and $v$ is the number of parameters in the maximal model not including the intercept. As we assume a second order maximal model, $v$ is the total number of main effects and two-factor interactions; $M_s(i, j)$ is an indicator for whether effects $i$ and $j$ are both in model $M_s$ and $\tilde{\operatorname{Pr}}\left(M_s\right)$ denotes the prior probability that the model $M_s$ is the model eventually chosen among all the eligible candidate models as in \cite{Factorscreening}.

The variance of the estimator $\operatorname{Var}\left(\hat{\beta}_{i}\right)$ under the centered parameterization in model $M_s$  is approximated by 
\begin{equation}\label{3.2}
 \operatorname{Var}\left(\hat{\beta}_{i}|M_s\right )\approx \sum_{j=0}^v \frac{a_{i j}{ }^2}{a_{i i}{ }^2 a_{j j}}M_s(i, j)=\sum_{j=0}^v r_{i j}M_s(i, j).
\end{equation}  
The approximation to the $A_s$ optimality criterion function (for all parameters except the intercept) for the model $M_s$ is then given by $
\sum_{i=1}^v \sum_{j=0}^v r_{i j} M_s(i, j)
$ under the centered parameterization as in \cite{MEAD}.

 This is an approximation to the full  $A_s$ optimality criterion expression obtained by replacing the exact variances of the parameter estimates with approximations. We wish to generalize the same idea to the baseline parameterization and it is clear that the prior probabilities of the models in equation (\ref{3.1}) will not change when we consider the baseline parameterization. Therefore it is sufficient to replace the approximation to the criterion function of $A_s$ optimality for the model $M_s$ under the centered parameterization by that under the baseline parameterization in order to obtain a neat form of $Q_B$ under the baseline parameterization. 

From the results in the last section we know that the estimator of an effect under the baseline parameterization can be written as a linear combination of estimators of effects under the centered parameterization. Therefore if we would like to do the approximation to the full $A_s$ optimality criterion expression under the baseline parameterization, the issue of approximating the covariances of estimators of effects under the centered parameterization will arise. Since there is no ideal solution to this issue in the literature, we will make the first-order assumption that the factorial effects are measured independently so that the covariances are zero.
 We can then deduce the approximation of the $A_s$ criterion function under the baseline parameterization for the model $M_s$ as 
\begin{equation}\label{3.3}
Q_B = \left(4 \sum_{i=1}^m \sum_{j=0}^v r_{i j}+24 \sum_{i=m+1}^v \sum_{j=0}^v r_{i j}\right)  M_s(i, j)
\end{equation}
if we have $m$ factors, and thus $m$ main effects, and $v$ is the total number of factorial effects excluding the intercept in the maximal model. 

In order to check whether our approximation of the $A_s$ criterion under the baseline parameterization works in practice, we have done some numerical checks on four designs assuming the model being used contains all the main effects and two-factor interactions. The designs consist of 4 factors and 12 runs and in particular, one of the designs considered is the minimal K-aberration design as in \cite{MukerjeeRahul2012}. The designs are shown in Table~\ref{tab:design-comparison-4-factor}.

\begin{table}[htbp]
\centering
\caption{Designs for 4 factors in 12 runs}
\label{tab:design-comparison-4-factor}
\resizebox{0.6\linewidth}{!}
{
\begin{tabular}{cc}
\renewcommand{\arraystretch}{0.7}
\begin{tabular}[t]{@{\hspace{8mm}}r@{\hspace{8mm}}r@{\hspace{8mm}}r@{\hspace{8mm}}r@{\hspace{8mm}}}
\hline
\( x_1 \) & \( x_2 \) & \( x_3 \) & \( x_4 \) \\
\hline
-1 & -1 & -1 & -1 \\
-1 & -1 & -1 & 1 \\
-1 & -1 & 1 & -1 \\
-1 & 1 & -1 & 1 \\
-1 & 1 & 1 & -1 \\
1 & 1 & 1 & 1 \\
-1 & -1 & -1 & 1 \\
1 & -1 & 1 & -1 \\
1 & -1 & 1 & 1 \\
1 & 1 & -1 & -1 \\
1 & 1 & -1 & 1 \\
1 & 1 & 1 & -1 \\
\hline
\multicolumn{4}{c}{Design 1} \\
\end{tabular}
& 
\renewcommand{\arraystretch}{0.7}
\begin{tabular}[t]{@{\hspace{8mm}}r@{\hspace{8mm}}r@{\hspace{8mm}}r@{\hspace{8mm}}r@{\hspace{8mm}}}
\hline
\( x_1 \) & \( x_2 \) & \( x_3 \) & \( x_4 \) \\
\hline
1 & 1 & -1 & 1 \\
-1 & 1 & 1 & 1 \\
-1 & -1 & -1 & -1 \\
1 & -1 & 1 & 1 \\
-1 & 1 & 1 & -1 \\
-1 & -1 & -1 & -1 \\
-1 & -1 & 1 & 1 \\
1 & -1 & 1 & -1 \\
1 & 1 & 1 & -1 \\
1 & 1 & -1 & -1 \\
-1 & 1 & -1 & 1 \\
1 & -1 & -1 & 1 \\
\hline
\multicolumn{4}{c}{minimal K-aberration design} \\
\end{tabular}
\\
\renewcommand{\arraystretch}{0.7}
\begin{tabular}[t]{@{\hspace{8mm}}r@{\hspace{8mm}}r@{\hspace{8mm}}r@{\hspace{8mm}}r@{\hspace{8mm}}}
\hline
\( x_1 \) & \( x_2 \) & \( x_3 \) & \( x_4 \) \\
\hline
-1 & -1 & -1 & -1 \\
-1 & -1 & -1 & 1 \\
-1 & -1 & 1 & -1 \\
-1 & 1 & -1 & 1 \\
-1 & 1 & 1 & -1 \\
-1 & 1 & 1 & 1 \\
1 & -1 & -1 & -1 \\
1 & -1 & 1 & 1 \\
1 & -1 & 1 & -1 \\
1 & 1 & -1 & 1 \\
1 & 1 & -1 & -1 \\
1 & 1 & 1 & 1 \\
\hline
\multicolumn{4}{c}{Design 2} \\
\end{tabular}
& 
\renewcommand{\arraystretch}{0.7}
\begin{tabular}[t]{@{\hspace{8mm}}r@{\hspace{8mm}}r@{\hspace{8mm}}r@{\hspace{8mm}}r@{\hspace{8mm}}}
\hline
\( x_1 \) & \( x_2 \) & \( x_3 \) & \( x_4 \) \\
\hline
-1 & -1 & -1 & -1 \\
-1 & -1 & -1 & 1 \\
-1 & -1 & -1 & -1 \\
-1 & -1 & 1 & 1 \\
-1 & -1 & 1 & -1 \\
-1 & -1 & 1 & 1 \\
1 & -1 & -1 & -1 \\
1 & -1 & 1 & 1 \\
1 & -1 & 1 & -1 \\
1 & 1 & -1 & 1 \\
1 & 1 & -1 & -1 \\
1 & 1 & 1 & 1 \\
\hline
\multicolumn{4}{c}{Design 3} \\
\end{tabular}
\end{tabular}
}
\end{table}

From Theorem~\ref{thm:theorem 1} we could equivalently state an $A_s$ optimization problem under the baseline parameterization as a weighted $A_s$ optimization problem under the centered parameterization and obtain the approximation based on that.
Thus, when considering the approximation of the baseline $A_s$ criterion we will stick to the ‘-1’ and ‘1’ scaling and when doing the exact $A_s$ criterion we switch the ‘-1’ to lower level ‘0’ in the baseline parameterization and the higher level ‘1’ stays the same.

For the exact $A_s$ optimality criterion in the baseline parameterization, we consider the minimization of the expression 
\begin{equation}\label{exact baseline $A_s$}
\operatorname{Var}\left(\hat{\theta}_1\right)+\cdots +\operatorname{Var}\left(\hat{\theta}_4\right)+\operatorname{Var}\left(\hat{\theta}_{12}\right)+\cdots +\operatorname{Var}\left(\hat{\theta_{34}}\right).
\end{equation}
Under this criterion, design 1 gives an $A_s$ value of 63; the minimal K-aberration design gives 23.67; design 2 gives 18.25; for design 3, the information matrix $X^{\top} X)$ is not invertible, indicating the $A_s$ optimality criterion tends to infinity. Therefore we might conclude that design 2 seems to be the optimal design among these four designs, followed by the minimal K-aberration design and design 1, and design 3 is the worst design. The approximated $A_s$ criterion of designs 1, 2, 3 and the minimal K-aberration design gives 18.44, 16.07, 20.53, 17.78 respectively. Since the rank order of these 4 designs are recovered by the approximation, we might conclude that our approximation of the $A_s$ criterion under the baseline parameterization, equation (\ref{3.3}), gives a reasonable ordering of designs.

The $Q_B$ expression under the baseline parameterization for the second order model is then defined as
\begin{equation}
\label{3.4}
Q_B = \sum_{s=1}^{n_0}\left(4 \sum_{i=1}^m \sum_{j=0}^v r_{i j}+24 \sum_{i=m+1}^v \sum_{j=0}^v r_{i j}\right) M_s(i, j) \tilde{\operatorname{Pr}}\left(M_s\right).
\end{equation}

The next step is to rearrange the expression in terms of generalized word counts as defined in \cite{Gilmour2010}. The equation above might be equivalently split up into$$
Q_B = 4 \sum_{s=1}^{n_0} \sum_{i=1}^m \sum_{j=0}^v r_{i j} M_s(i, j)\tilde{\operatorname{Pr}} \left(M_s\right)+24 \sum_{s=1}^{n_0 } \sum_{i=m+1}^v \sum_{j=0}^v r_{i j} M_s(i, j)\tilde{\operatorname{Pr}}\left(M_s\right).
$$

Since $p_{i j}=\sum_{s=1}^{n_0} \tilde{\operatorname{Pr}}\left(\mathcal{M}_s\right) M_s(i, j)$ as in \cite{Factorscreening}, we might rewrite the above expression as 
$$
Q_B = 4 \sum_{i=1}^m \sum_{j=0}^v r_{i j} p_{i j}+24 \sum_{i=m+1}^v \sum_{j=0}^v r_{i j} p_{i j}.
$$
Following the notation in \cite{Gilmour2010}, we further rewrite our expression for $Q_B$ under the baseline parameterization for second order maximal model as 
\begin{equation}
\label{3.5}
Q_B = 4 \sum_{i=1}^m \sum_{j=0}^v p_{i j} \frac{a_{i j}{ }^2}{n^2}+24 \sum_{i=m+1}^v \sum_{j=0}^v p_{i j} \frac{a_{i j}{ }^2}{n^2}.
\end{equation}

We now split our equation in the same way as discussed in Appendix $B.2$ in \cite{Gilmour2010} into 5 groups as 
\begin{equation}
\label{3.6}
\begin{split}
Q_B = 4 \sum_{i=1}^m p_{i 0} \frac{a_{i 0}{ }^2}{n^2}+4 \sum_{i=1}^m \sum_{\substack{j=1 \\ j \neq i}}^m p_{i j} \frac{a_{i j}{ }^2}{n^2}+4 \sum_{i=1}^m \sum_{\substack{j=m+1 \\ j \neq i}}^v p_{i j} \frac{a_{i j}{ }^2}{n^2}+\\24 \sum_{i=m+1}^v p_{i 0} \frac{a_{i 0}{ }^2}{n^2}+24 \sum_{i=m+1}^v \sum_{\substack{j=m+1 \\ j \neq i}}^v p_{i j} \frac{a_{i j}{ }^2}{n^2}+
24 \sum_{i=m+1}^v \sum_{\substack{j=1 \\ j \neq i}}^m p_{i j} \frac{a_{i j}{ }^2}{n^2}.
\end{split}
\end{equation}
 The first term of equation (\ref{3.6}) shall be equivalent to $i$ referring to the intercept and $j$ referring to the main effect of a factor and corresponds to group 1 in the appendix $B.2$ in \cite{Gilmour2010}; the second term corresponds to the case when $i,j$ refer to main effects of a pair of factors in group 3;  the fourth term shall be equivalent to $i$ referring to the intercept and $j$ referring to a two-factor interaction discussed in group 2 and the fifth term contains two cases in group 4 where $i,j$ refer to a pair of interactions.

Group 5 in appendix $B.2$ in \cite{Gilmour2010} is where more careful consideration is needed.
When $i$ is a main effect and $j$ is an interaction the sum of $\frac{a_{i j}{ }^2}{n^2}$ is $(m-1) b_1(1)$ when they have one factor in common and $3 b_3(3)$ otherwise and the same is true for the reverse case. The terms $\sum_{i=1}^m \sum_{\substack{j=m+1 \\ j \neq i}}^v p_{i j} \frac{a_{i j}{ }^2}{n^2}$ and $\sum_{i=m+1}^v\sum_{\substack{j=1 \\ j \neq i}}^m  p_{i j} \frac{a_{i j}{ }^2}{n^2}$ have both been assigned weight 1 in \cite{Gilmour2010} and thus the sum of $\frac{a_{i j}{ }^2}{n^2}$ is $2(m-1) b_1(1)$ for $i$ being a main effect and $j$ being and interaction or vice versa and $6 b_3(3)$ otherwise. In equation (\ref{3.6}), these two terms are assigned different weights so it shall be more reasonable to treat them separately as $4\left(\xi_{21} (m-1) b_1(1)+\xi_{31} 3 b_3(3)\right)$+$24\left(\xi_{21} (m-1) b_1(1)+\xi_{31} 3 b_3(3)\right)$.

 Thus following the same argument in the appendix $B.2$ in \cite{Gilmour2010}, we obtain a version of $Q_B$ expression for evaluating the designs under the baseline parameterization up to the proper definition of $\xi_{ab}$ and $ b_i(i)$ and simplifications as
\begin{equation}
\label{3.8}
\begin{split}
Q_B = 4 \xi_{10} b_1(1)+8 \xi_{20} b_2(2)+28\left(\xi_{21} (m-1) b_1(1)+\xi_{31} 3 b_3(3)\right)\\+24 \xi_{21} b_2(2)+24\left(\xi_{32} 2(m-2) b_2(2)+\xi_{42} 6 b_4(4)\right).
\end{split}
\end{equation}
We first give the definition of the set of prior probabilities $\pi_1$ and $\pi_2$. $\pi_1$ denotes the prior belief that any linear term $x_i$ is in the model and $\pi_2$ denotes the prior probability that any interaction is in the true model given that the linear effects of both factors involved are in the model. Since by this definition we have assumed the prior probability for each effect of the same type in the best model is the same. The prior probability for a model to be chosen depends only on the number of main effects and two-factor interactions in this model. More specifically, the prior probability for a model with exactly $a$ main effects and $b$ two-factor interactions is given by 
\begin{equation}
\label{3.102}
\operatorname{Pr}\left(M_{s}\right)=\pi_{1}^{a}\left(1-\pi_{1}\right)^{m-a}  \pi_{2}^{b}\left(1-\pi_{2}\right)^{\frac{a(a-1)}{2}-b},
\end{equation} as in \citet{TG2024}. $\xi_{ab}$ is then defined as the sum of the prior probabilities of models containing a specific set of $a$ main effects and a particular set of $b$ two-factor interactions out of these $a$ main effects and following this definition the expression for $\xi_{ab}$ can be summarized as $$\xi_{ab}=\pi_1^a \pi_2^b,$$ 
where the detailed derivation could be found in \citet{TG2024}.

The generalized word count, $ b_i(i)$, does the summation of the aliasing between the intercept and individual effects of order $i\geq 1$; its definition can be found in \citet{Gilmour2010}. It can be computed by
\begin{equation}
\label{3.101}
b_i(i)=\sum_{|s|=i} R_i(s) ; \quad R_i(s)=\frac{1}{N^2}\left[\sum_{h=1}^N\left(X_{h 1} \cdots X_{h i}\right)\right]^2,
\end{equation}
where $X_1, \cdots, X_m$ are all the columns forming the design and $s$ is a subset of $i$ columns of this full set.

After rearranging we shall obtain our $Q_B$ expression under the baseline parameterization for the second order maximal model as
\begin{equation}
\label{3.9}
\begin{split}
Q_B = \left\{4 \xi_{10}+28(m-1) \xi_{21}\right\} b_1(1)+\left\{8 \xi_{20}+24 \xi_{21}+48(m-2)\xi_{32}\right\} b_2(2)+\\ 84\xi_{31}b_3(3)+144\xi_{42}b_4(4).
\end{split}
\end{equation}

When the above expression is minimised, so is $\frac{1}{4}$ of it and therefore the expression we are going to use is 
\begin{equation}
\label{3.100}
\begin{split}
Q_B \propto \left\{ \xi_{10}+7(m-1) \xi_{21}\right\} b_1(1)+\left\{2 \xi_{20}+6 \xi_{21}+12(m-2)\xi_{32}\right\} b_2(2)+\\ 21\xi_{31}b_3(3)+36\xi_{42}b_4(4).
\end{split}
\end{equation}
We might notice that the $Q_B$ expression we obtain here that is applicable to the baseline parameterization shares the same format as its counterpart in \citet{Gilmour2010}, but with different coefficients. This gives the implication that in general the set of $Q_B$ optimal designs found under the centered parameterization shall not be the same set of $Q_B$ optimal designs suitable for the baseline parameterization.

We have seen how $\xi_{ab}$ is associated with the prior probabilities $\pi_1$ and $\pi_2$ in the previous sections. This might be useful in the sense that compared with other optimality criteria in the literature like \cite{MukerjeeRahul2012}, one design might be sub-optimal compared with another design under one set of prior beliefs of $\pi_1$ and $\pi_2$, but it might then perform better for another set of prior beliefs of $\pi_1$ and $\pi_2$, which will depend on practical considerations. This might give us additional flexibility compared with the criterion developed in \cite{MukerjeeRahul2012}.

\section{Evaluating designs in the literature}
\label{sec:evaluation}

From the previous section we might verify that the minimum K-aberration design will give a $Q_B$ value of 0, which is the minimum, when $\pi_2=0$. This suggests the minimum K-aberration design is optimal when the main effects maximal model is assumed. Under the current framework, we could vary the prior probabilities $\pi_1$ and $\pi_2$ to reflect experimenters' opinion on the possible best fitting model going beyond the main effects model. A set of new optimal designs will arise, which need to be obtained by searching, in addition to the minimum K-aberration design. We will illustrate this idea in the following example.

Consider the four design matrices in Table ~\ref{tab:design-comparison}, where the top left corresponds to a minimum K-aberration design with 6 factors and 12 runs given by \cite{MukerjeeRahul2012}. In order to operate under 
our framework, we rewrite the design matrix under the centered parameterization shown in the bottom left. Two alternative designs are shown on the right. We call the designs in the top and bottom right $AD_1$ and $AD_2$ respectively. We are now ready to present some numerical results.

\begin{table}[htbp]
\centering
\caption{The minimal K-aberration design and alternatives}
\label{tab:design-comparison}
\resizebox{0.8\linewidth}{!}
{
\begin{tabular}{cc}

\renewcommand{\arraystretch}{0.7}
\begin{tabular}[t]{@{\hspace{5mm}}r@{\hspace{5mm}}r@{\hspace{5mm}}r@{\hspace{5mm}}r@{\hspace{5mm}}r@{\hspace{5mm}}r@{\hspace{5mm}}}
\\
0 & 0 & 0 & 0 & 0 & 0 \\
0 & 0 & 0 & 0 & 0 & 1 \\
1 & 1 & 0 & 0 & 1 & 0 \\
1 & 0 & 1 & 0 & 1 & 1 \\
0 & 1 & 0 & 1 & 1 & 1 \\
1 & 1 & 1 & 0 & 0 & 1 \\
1 & 0 & 1 & 1 & 0 & 0 \\
1 & 1 & 0 & 1 & 0 & 0 \\
0 & 1 & 1 & 0 & 1 & 0 \\
0 & 1 & 1 & 1 & 0 & 1 \\
0 & 0 & 1 & 1 & 1 & 0 \\
1 & 0 & 0 & 1 & 1 & 1 \\
\\
\multicolumn{6}{c}{Min K} \\
\end{tabular}
& 
\renewcommand{\arraystretch}{0.7}
\begin{tabular}[t]{@{\hspace{5mm}}r@{\hspace{5mm}}r@{\hspace{5mm}}r@{\hspace{5mm}}r@{\hspace{5mm}}r@{\hspace{5mm}}r@{\hspace{5mm}}}
\\
-1 & 1 & -1 & 1 & 1 & -1 \\
-1 & -1 & -1 & 1 & 1 & 1 \\
-1 & -1 & -1 & -1 & -1 & 1 \\
1 & 1 & -1 & -1 & 1 & 1 \\
1 & 1 & -1 & 1 & -1 & 1 \\
1 & 1 & 1 & 1 & 1 & -1 \\
1 & -1 & 1 & 1 & 1 & 1 \\
-1 & 1 & -1 & -1 & -1 & -1 \\
1 & -1 & 1 & -1 & -1 & 1 \\
1 & 1 & 1 & -1 & -1 & -1 \\
-1 & -1 & 1 & 1 & -1 & -1 \\
-1 & -1 & 1 & -1 & 1 & -1 \\
\\
\multicolumn{6}{c}{$AD_1$} \\
\end{tabular}
\\
\renewcommand{\arraystretch}{0.7}
\begin{tabular}[t]{@{\hspace{5mm}}r@{\hspace{5mm}}r@{\hspace{5mm}}r@{\hspace{5mm}}r@{\hspace{5mm}}r@{\hspace{5mm}}r@{\hspace{5mm}}}
\\
-1 & -1 & -1 & -1 & -1 & -1 \\
-1 & -1 & -1 & -1 & -1 & 1 \\
1 & 1 & -1 & -1 & 1 & -1 \\
1 & -1 & 1 & -1 & 1 & 1 \\
-1 & 1 & -1 & 1 & 1 & 1 \\
1 & 1 & 1 & -1 & -1 & 1 \\
1 & -1 & 1 & 1 & -1 & -1 \\
1 & 1 & -1 & 1 & -1 & -1 \\
-1 & 1 & 1 & -1 & 1 & -1 \\
-1 & 1 & 1 & 1 & -1 & 1 \\
-1 & -1 & 1 & 1 & 1 & -1 \\
1 & -1 & -1 & 1 & 1 & 1 \\
\\
\multicolumn{6}{c}{Min K Centered} \\
\end{tabular}
& 
\renewcommand{\arraystretch}{0.7}
\begin{tabular}[t]{@{\hspace{5mm}}r@{\hspace{5mm}}r@{\hspace{5mm}}r@{\hspace{5mm}}r@{\hspace{5mm}}r@{\hspace{5mm}}r@{\hspace{5mm}}}
\\
-1 & 1 & 1 & -1 & 1 & -1 \\
-1 & -1 & -1 & -1 & 1 & -1 \\
-1 & 1 & 1 & -1 & -1 & 1 \\
-1 & 1 & -1 & 1 & -1 & -1 \\
-1 & -1 & -1 & 1 & -1 & 1 \\
-1 & -1 & 1 & 1 & 1 & 1 \\
1 & -1 & 1 & 1 & -1 & -1 \\
1 & -1 & 1 & -1 & 1 & -1 \\
1 & 1 & 1 & 1 & 1 & -1 \\
1 & -1 & -1 & -1 & -1 & 1 \\
1 & 1 & -1 & -1 & 1 & 1 \\
1 & 1 & -1 & 1 & -1 & 1 \\
\\
\multicolumn{6}{c}{$AD_2$} \\
\end{tabular}
\end{tabular}
}
\end{table}

 From the genralized word counts in Table~\ref{tab:bii} we calculate the  $Q_B$ value for these three designs under various pairs of prior probabilities $\pi_1$ and $\pi_2$ giving the results in Table~\ref{QBK3}.
\begin{table}[htbp] 
\caption{The generalized word count pattern for min K-aberration, $AD_1$ and $AD_2$ }
\label{tab:bii}
\centering  
{
\begin{tabular}{cclll} 
\hline
& $b_1(1)$ & $b_2(2)$ & $b_3(3)$ & $b_4(4)$ \\
\hline 
\text{Min K} & 0 & 0 & 2.2222 & 1.6667 \\
\hline 
$AD_1$ & 0 & 0.7778 & 0 & 3.4444 \\
\hline 
$AD_2$ & 0 & 0.4444 & 1.5556 & 1.2222 \\
\hline

\end{tabular}
}
\end{table}

\begin{table}[htbp] 
\caption{$Q_B$ values for min K-aberration design, $AD_1$ and $AD_2$}
\label{QBK3}
\centering
\resizebox{0.55\linewidth}{!}{
\begin{tabular}{lrrr} 
\hline ($\pi_1,\pi_2$) & min K  & $A D_1$ & $A D_2$ \\
\hline(0.4,0.2) & 0.6588 & 0.6208 & 0.7454 \\
\hline(0.6,0.4) & 5.2762 & 5.0935 & 5.1761 \\
\hline(0.6,0.6) & 8.8474 & 10.2564 & 8.8413 \\
\hline(0.8,0.4) & 13.4895 & 13.3750 & 12.5729 \\
\hline(0.8,0.6) & 23.1834 & 27.9534 & 22.0483 \\
\hline

\end{tabular}
}
\end{table}

From the numerical results we might conclude that, the minimal K-aberration design does not give the smallest $Q_B$ value, and this indicates that the $Q_B$ optimal design at these priors is different from the minimal K-aberration design and must be found by other means.

\begin{figure}[htbp]
\centering
\includegraphics [angle=0, scale=0.6]{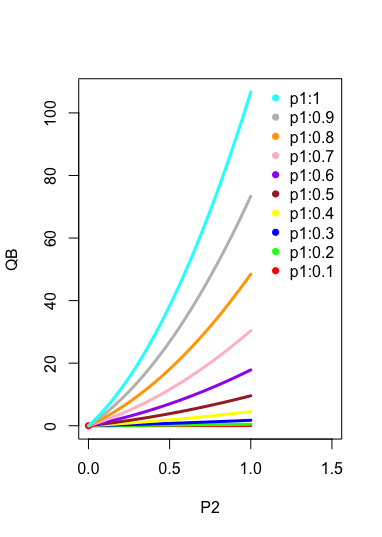}\par
\caption{min K-aberration design at various priors}
\label{fig:figure 1}
\end{figure}

From Figure~\ref{fig:figure 1} we see that at one given $\pi_1$, the $Q_B$ value for the minimal K-aberration design approaches 0 as $\pi_2$ converges to 0 from above. Comparing the $Q_B$ curves from various $\pi_1$, it is also clear that as $\pi_1$ increases, the maximum $Q_B$ value increases dramatically, as does the rate of curves to reach its maximum $Q_B$ value. For instance, for $\pi_1$ up to 0.2, the rate of growth is very slow and their maximum $Q_B$ is close to 0. This might suggest that for $\pi_1$ up to 0.2 and all the possible $\pi_2$, the minimal K-aberration design might be optimal, while this might not be the case when $\pi_1=1$

\section{Finding optimal designs}
\label{sec:optimal}
We have understood the behavior of the minimal K-aberration design, and thus we set out to find $Q_B$ optimal designs using an extended version of the coordinate exchange algorithm which will be explained shortly. We will consider two cases for illustration: 6 factors with 12 runs and 9 factors with 16 runs. 

A case study was given in \cite{Goos} relating to a chemical substance extraction experiment consisting of 6 factors and with a budget sufficient for 12 runs. The design considered has the same $Q_B$ value as the minimal K-aberration design, and thus we search for other optimal designs separately in our case. 

It has sometimes been found that coordinate exchange can find better designs using ``wrong'' priors than using the actual priors specified, e.g.\  \citet{egorova2023optimalresponsesurfacedesigns}. Our extended coordinate exchange algorithm exploits this by running for multiple priors. It consists of three steps. 
 \begin{enumerate}
     \item Conduct the traditional coordinate exchange. More specifically, we randomly generate the starting design and conduct the coordinate exchange. If we have a design of $m$ factors and $n$ runs, we consider the $m\times n$ coordinates in a single cycle. For each coordinate, we switch it from 1 to -1 or vice versa and then calculate the $Q_B$ value. If that switch lowers the value of $Q_B$ we retain this switch, otherwise we switch it back. The improved design in one cycle will then be the starting design for the next cycle and we continue doing this until no improvement is possible, and we then restart from a new starting design. When the pre-specified number of starting designs have been used we consider the traditional approach of coordinate exchange is finished and we move to Step 2. 

\item For each prior pair $\pi_{1i}$ and $\pi_{2i}$, calculate the $Q_B$ efficiency of the optimal design at the prior $\pi_{1i}$ and $\pi_{2i}$ versus all the optimal designs obtained at $\pi_{1j}$ and $\pi_{2j}$ differing from $\pi_{1i}$ and $\pi_{2i}$. That is, we will calculate the quantity $\frac{Q_B(OD(\pi_{1i},\pi_{2i}))}{Q_B(OD(\pi_{1j},\pi_{2j}))}$ where $OD(\pi_{1i},\pi_{2i})$ refers to the optimal design obtained at the prior $\pi_{1i}$ and $\pi_{2i}$.

 \item The quantity in Step 2 is usually such that $\frac{Q_B(OD(\pi_{1i},\pi_{2i}))}{Q_B(OD(\pi_{1j},\pi_{2j}))}\leq 1$. If that is not the case, we will replace $OD(\pi_{1i},\pi_{2i})$ by $OD(\pi_{1j},\pi_{2j})$, where the latter design is used as a starting design and we conduct Step 1 on this starting design to obtain the optimal design.
\end{enumerate}

We finish the discussion of this subsection by noting that since our extended coordinate algorithm allows the comparison of $Q_B$ optimal designs across different prior pairs, it has sometimes avoided the issue of finding local minima compared to traditional coordinate exchange algorithm. 

\subsection{Results for 6 factors and 12 runs}

We first present the result for the priors $\pi_1$ and $\pi_2$ both taking values in $\{0.2,0.4,0.6,0.8,1\}$. Eleven optimal designs were found, labelled $D_1$ to $D_{11}$, which have the generalized word counts shown in Table~\ref{tab:biio}.

\begin{table}[htbp]  
\caption{Generalized word counts for the optimal designs}
\label{tab:biio}
\centering  
\resizebox{0.5\linewidth}{!}{
\begin{tabular}{ccccc} 
\hline
&$b_1(1 )$& $b_2(2 )$ & $b_3(3 )$ & $b_4(4 )$ \\
\hline
D1 & 0 & 0 & 2.2222 & 1.6667  \\
D2 & 0 & 0.6667 & 0 & 3.6667  \\
D3 & 0 & 0.1111 & 1.7778 & 1.8889  \\
D4 & 0 & 0.3333 & 0.8889 & 3  \\
D5 & 0 & 0.2222 & 1.3333 & 2.3333  \\
D6 & 0 & 0.2222 & 1.7778 & 1.4444  \\
D7 & 0 & 1.5556 & 0.4444 & 0.1111  \\
D8 & 0.0278 & 0.1389 & 1.8333 & 1.5  \\
D9 & 1.2222 & 0.3333 & 0.2222 & 0.3333  \\
D10 & 0.0278 & 0.1389 & 2.5 & 0.8333  \\
D11 & 0.0556 & 0.3333 & 2.5556 & 0.3333  \\
\hline

\end{tabular}
}
\end{table}

We note here that design $D_1$ has the same $Q_B$ value as the minimal K-aberration design for all $\pi_1$ and $\pi_2$. For each prior, we have found the $Q_B$ optimal design and compared the value of $Q_B$ with the $Q_B$ value obtained from the minimal K-aberration design given by \cite{MukerjeeRahul2012}, and thus we have obtained the relative $Q_B$-efficiency of the minimal K-aberration design as shown in Table ~\ref{tab:re}.

\begin{table}
\caption{Relative $Q_B$-efficiency of the minimal K-aberration design}
\label{tab:re}
\begin{center}
\resizebox{0.55\linewidth}{!}{
\begin{tabular}{cccccc} 
\hline
$\pi_1$ & $\pi_2$ & $Q_B$ for min K & best $Q_B$ & $Q_B$ efficiency & Design \\
\hline
0.2 & 0.2 & 0.0785 & 0.0785 & 1.0000 & D1 \\
0.2 & 0.4 & 0.1647 & 0.1633 & 0.9913 & D3 \\
0.2 & 0.6 & 0.2586 & 0.2586 & 1.0000 & D1 \\
0.2 & 0.8 & 0.3601 & 0.3601 & 1.0000 & D1 \\
0.2 & 1 & 0.4693 & 0.4693 & 1.0000 & D1 \\
0.4 & 0.2 & 0.6588 & 0.5584 & 0.8477 & D2 \\
0.4 & 0.4 & 1.4404 & 1.3187 & 0.9155 & D4 \\
0.4 & 0.6 & 2.3450 & 2.2827 & 0.9735 & D3 \\
0.4 & 0.8 & 3.3724 & 3.3649 & 0.9978 & D3 \\
0.4 & 1 & 4.5227 & 4.5227 & 1.0000 & D1 \\
0.6 & 0.2 & 2.3270 & 1.7288 & 0.7429 & D2 \\
0.6 & 0.4 & 5.2762 & 4.8817 & 0.9252 & D5 \\
0.6 & 0.6 & 8.8474 & 8.5341 & 0.9646 & D8 \\
0.6 & 0.8 & 13.0406 & 12.6900 & 0.9731 & D8 \\
0.6 & 1 & 17.8560 & 17.4347 & 0.9764 & D10 \\
0.8 & 0.2 & 5.7617 & 4.1834 & 0.7261 & D2 \\
0.8 & 0.4 & 13.4895 & 12.5533 & 0.9306 & D6 \\
0.8 & 0.6 & 23.1834 & 21.8990 & 0.9446 & D6 \\
0.8 & 0.8 & 34.8433 & 32.6773 & 0.9378 & D10 \\
0.8 & 1 & 48.4693 & 43.5801 & 0.8991 & D11 \\
1 & 0.2 & 11.7333 & 8.6933 & 0.7409 & D2 \\
1 & 0.4 & 28.2667 & 23.1644 & 0.8195 & D7 \\
1 & 0.6 & 49.6000 & 41.6356 & 0.8394 & D9 \\
1 & 0.8 & 75.7333 & 59.3644 & 0.7839 & D9 \\
1 & 1 & 106.6667 & 79.3333 & 0.7438 & D9 \\
\hline

\end{tabular}
}
\end{center}
\end{table}


Given the set of optimal designs as in Table ~\ref{tab:biio}, it is interesting to explore the range of prior probabilities $\pi_1$ and $\pi_2$ for which a particular design in Table ~\ref{tab:biio} is $Q_B$ optimal versus all other designs in Table ~\ref{tab:biio}. This shall be particularly useful when recommending designs to experimenters based on their choice of priors. For any given design we first solve the $Q_B$ equation versus all competing designs to find the contour where they perform equally well. Based on these contours we work out the region where the targeted design is better in $Q_B$ versus the other design. Repeating this process for all competing designs and finding the joint intersection we will be able to locate the area where one given design is $Q_B$ optimal against all other competing designs. 

As an example for finding a contour we consider designs $D_2$ and $D_4$ in Table~\ref{tab:biio}. To establish and solve the equation we substitute the $b_i(i )$ in equation (\ref{3.100}) with the $b_i(i )$ from designs $D_2$ and $D_4$, replace $\xi_{ab}$ with $\pi_1^a\pi_2^b $ and equate these two equations. The equation to be solved is
\begin{equation}
1+\left(3-28 \pi_1\right) \pi_2+\left(24 \pi_1+36 \pi_1^2\right) \pi_2^2=0
\end{equation}
and the well-defined solution is given by
\begin{equation}
\pi_2 = \frac{28 \pi_1 - 3 \pm \sqrt{9 - 264 \pi_1 + 640 \pi_1^2}}{8 \pi_1 \left(6 + 9 \pi_1\right)}
\end{equation}
for $0 \leq\pi_1, \pi_2 \leq 1$ from the above expression.

\begin{figure}[htbp]
\centering
\includegraphics [angle=0, scale=0.6]{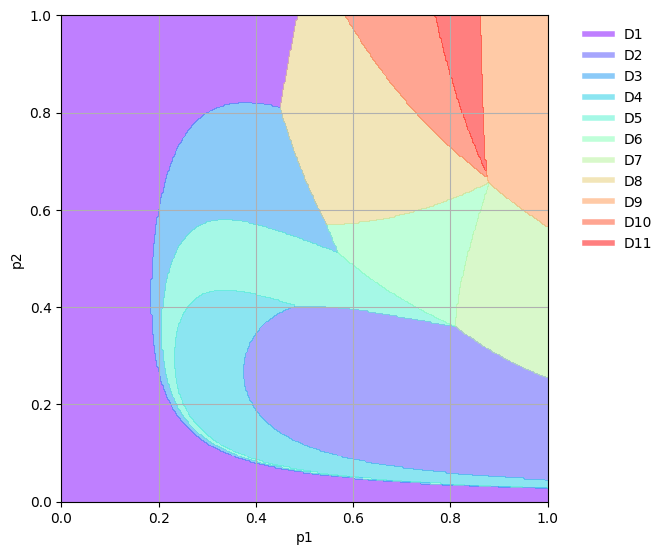}\par
\caption{Regions for each individual design to be $Q_B$ optimal}
\label{fig:figure 2}
\end{figure}

We now present the graph of regions for each design in Table~\ref{tab:biio} to be universally $Q_B$ optimal against all other designs in Figure~\ref{fig:figure 2}. From this figure we can tell that the optimal design depends on the prior in a complicated manner. Also, it's impossible to estimate all the parameters for submodels that have number of parameters larger than the size of experiment. The number of parameters in a submodel can be defined through the number of factors and prior probabilities $\pi_1$ and $\pi_2$. We will see detailed discussion regarding to this in Section~\ref{sec:Eva}.

\clearpage

\subsection{Results for 9 factors and 16 runs}

We now present the results for 9 factors in 16 runs, with the prior probabilities $\pi_1$ and $\pi_2$ from the set $\{0.1,0.3,0.5,0.7,0.9\}$. We use a different set of prior probabilities for this example compared to the first example for illustration. One may also explore other possible prior probabilities. The optimal designs in this case have the generalized word counts given in Table~\ref{tab:916}.

\begin{table}[htbp]  
\caption{Generalized word counts of optimal designs for 9 factors in 16 runs}
\label{tab:916}
\begin{center}
\resizebox{0.5\linewidth}{!}{
\begin{tabular}{ccccc} 
\hline
 &$b_1(1 )$& $b_2(2 )$ & $b_3(3 )$ & $b_4(4 )$ \\
\hline
D1 & 0 & 0 & 4 & 14 \\
D2 & 0 & 0 & 6 & 9 \\
D3 & 0 & 1 & 0 & 21 \\
D4 & 0.0938 & 0.9063 & 2.7188 & 11.2813 \\
$D_L$& 0 & 1 & 3 & 11 \\
D5 & 0.1875 & 0.8125 & 2.4375 & 11.5625 \\
\hline
\end{tabular}
}
\end{center}
\end{table}

The best $Q_B$ values obtained and the relative efficiencies of the minimal K-aberration design against the $Q_B$ optimal designs is summarised in Table~\ref{tab:916effi}. D4 from Table~\ref{tab:916} which is found to be $Q_B$ optimal at $\pi_1$=0.7, $\pi_2$=0.3 by extended coordinate exchange has been updated to $D_L$ in Table~\ref{tab:916effi} which is found by conducting optimization restricted to level balanced designs and has a lower $Q_B$ value.

\begin{table}[htbp]  
\caption{Relative efficiency of the minimal K-aberration design against the optimal designs found}
\label{tab:916effi}
\begin{center}
\resizebox{0.55\linewidth}{!}{
\begin{tabular}{cccccc} 
\hline
$\pi_1$ & $\pi_2$ & $Q_B$ for min K & best $Q_B$ & $Q_B$ efficiency & Design \\
\hline
0.1 & 0.1 & 0.0089 & 0.0089 & 1 & D1 \\
0.1 & 0.3 & 0.0297 & 0.0297 & 1 & D1 \\
0.1 & 0.5 & 0.0546 & 0.0546 & 1 & D1 \\
0.1 & 0.7 & 0.0835 & 0.0835 & 1 & D1 \\
0.1 & 0.9 & 0.1164 & 0.1164 & 1 & D1 \\
0.3 & 0.1 & 0.2676 & 0.2676 & 1 & D1 \\
0.3 & 0.3 & 1.0478 & 1.0478 & 1 & D1 \\
0.3 & 0.5 & 2.1546 & 2.1546 & 1 & D1 \\
0.3 & 0.7 & 3.5880 & 3.5880 & 1 & D1 \\
0.3 & 0.9 & 5.3479 & 5.1876 & 0.9700 & D2 \\
0.5 & 0.1 & 1.3650 & 1.2275 & 0.8993 & D3 \\
0.5 & 0.3 & 5.9850 & 5.9850 & 1 & D1 \\
0.5 & 0.5 & 13.1250 & 12.9375 & 0.9857 & D2 \\
0.5 & 0.7 & 22.7850 & 20.9475 & 0.9194 & D2 \\
0.5 & 0.9 & 34.9650 & 30.5775 & 0.8745 & D2 \\
0.7 & 0.1 & 4.0913 & 3.3773 & 0.8255 & D3 \\
0.7 & 0.3 & 19.5345 & 19.4949 & 0.9980 & $D_L$ \\
0.7 & 0.5 & 44.6586 & 41.0571 & 0.9194 & D2 \\
0.7 & 0.7 & 79.4635 & 68.3709 & 0.8604 & D2 \\
0.7 & 0.9 & 123.9492 & 101.9080 & 0.8222 & D2 \\
0.9 & 0.1 & 9.4303 & 7.6785 & 0.8142 & D3 \\
0.9 & 0.3 & 48.1315 & 45.4729 & 0.9448 & D5 \\
0.9 & 0.5 & 113.2866 & 99.0711 & 0.8745 & D2 \\
0.9 & 0.7 & 204.8957 & 168.4602 & 0.8222 & D2 \\
0.9 & 0.9 & 322.9587 & 254.8555 & 0.7891 & D2 \\
\hline
\end{tabular}
}
\end{center}
\end{table}

\begin{figure}[htbp]
\centering
\includegraphics [angle=0, scale=0.6]{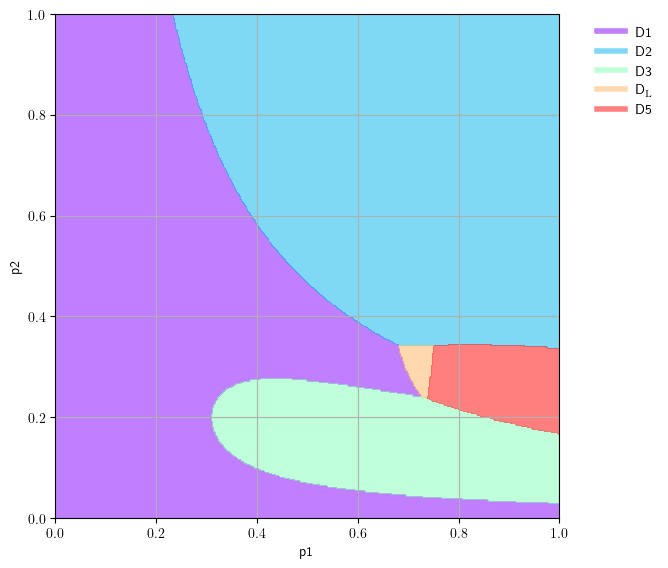}\par
\caption{Regions for each individual design to be $Q_B$ optimal}
\label{fig:figure 3}
\end{figure}

Like the case of 6 factors in 12 runs, we obtain a graph, as Figure~\ref{fig:figure 3}, for the optimal designs in Table~\ref{tab:916} such that each one of them is $Q_B$ optimal against other designs for a range of priors. This graph shows that $Q_B$ optimal designs at most, but not all, prior probabilities are given by the orthogonal designs $D_1$ and $D_2$.

\subsection{Evaluating the optimal designs}
\label{sec:Eva}
Since we have obtained the $Q_B$ optimal designs, we wish to assess their performances given the expected fitted models versus the reference minimal K-aberration design to illustrate the usefulness of $Q_B$ optimal designs.

As the $Q_B$ criterion is an approximation of the average $A_s$ optimality criterion over many models, we calculated the $A_s$ criterion function values of both designs for different fitted models of typical sizes from the corresponding prior probabilities under the baseline parameterization. For each size of fitted model, we denote the cardinality of all the eligible models by $n_T$ and consider two criteria to evaluate the designs.

Before going further, we note that not all pairs of prior probabilities are being evaluated here, because the current framework deals with estimable models only, where estimable models are those models with the number of parameters at most the run size of the experiment. The expected number of main effects included in the model is determined by rounding the expression $M=m\pi_1$ to the nearest integer and the expected number of interactions in the model is given by rounding the expression $N_I=\binom{M}{2} \pi_2$ to the nearest integer. $n_T$ is then given by the expression 
$$
n_T = \binom{m}{M}\times\binom{\binom{M}{2}}{N_I}
$$

First of all, the $Q_B$ criterion looks for designs that give good parameter estimation in as many models as possible within a specified maximal model. Following this aim we define $\frac{n(Q_B)}{n_T}$ and $\frac{n(K)}{n_T}$ where $n(Q_B)$ and $n(K)$ are number of models out of $n_T$ models for which a finite $A_s$ value can be calculated from the $Q_B$ optimal design and minimal K-aberration design respectively. The metrics $\frac{n(Q_B)}{n_T}$ and $\frac{n(K)}{n_T}$ will then give the ratio of submodels that could be estimated by these two types of designs.

This idea is similar to the estimation capacity in \cite{10.1111/1467-9868.00164}
and projection estimation capacity considered in \cite{LI2008154}. While estimation capacity considers submodels with all the main effects and a selected number of two-factor interactions, projection capacity considers submodels with a subset of main effects and all the two-factor interactions from these main effects. Compared to these criteria, our metrics could be more general and flexible. In column 3 and column 4 in Table~\ref{tab:left_table} and Table~\ref{tab:right_table} we found that in all but three cases, the $Q_B$ optimal design is not inferior to or better than minimal-K aberration design in terms of projection properties. This implies that in most cases The $Q_B$ optimal design has the same or better projection efficiency compared to the minimal K-aberration design. For the last three sizes of fitted models in Table~\ref{tab:right_table} the $Q_B$ optimal design has achieved overwhelming advantages in terms of the projection efficiency. 

We have also calculated the average $A_s$ value given by the $Q_B$ optimal design and the minimal-K aberration design for the submodels estimable under both designs denoted by $\frac{\sum_{Q_B}A_s}{n(Q_B\land K)}$ and $\frac{\sum_{K}A_s}{n(Q_B\land K)}$ respectively. For columns 6 and 7 in both tables we found in most cases the $Q_B$ optimal design is slightly inferior in terms of average $A_s$ value for 9 factors and has achieved better average $A_s$ value for 6 factors. At $\pi_1=0.8$ and $\pi_2=0.6$ for 6 factors, the average $A_s$ of the $Q_B$ optimal design is almost twice of that for the minimal K-aberration, but 139 more submodels could be estimated from the $Q_B$ optimal design than from the minimal K-aberration design. Analogous situations happen at fitted models corresponding to the priors (0.7,0.3) and (0.9,0.1) for 9 factors. The number of submodels that could be estimated by the $Q_B$ optimal design is nearly 3 times that under the minimal K-aberration design.   

It is worth noting that for 9 factors and the prior pair (0.7,0.3), Table~\ref{tab:right_table} gives the projection performance of design $D_L$. If one wishes to obtain a design with an even better projection efficiency, that's possible since design $D_4$ from Table~\ref{tab:916} has the projection effiency $\frac{n(Q_B)}{n_T}=0.98$ and the trade off to be made is a slightly higher $Q_B$ value 19.5257 compared to that in Table~\ref{tab:916effi} and a slightly higher average $A_s$ value of 16.44.


\begin{table}[htbp]
\begin{minipage}[t]{0.48\linewidth}
\centering
\caption{$Q_B$ vs min K in 6 factor fitted models}
\label{tab:left_table}
\resizebox{\linewidth}{!}{
\begin{tabular}{ccccccc} 
\hline
$\pi_1$ & $\pi_2$ & $\frac{n(Q_B)}{n_T}$ & $\frac{n(K)}{n_T}$ & $n_T$  & $\frac{\sum_{Q_B}A_s}{n(Q_B\land K)}$&$\frac{\sum_{K}A_s}{n(Q_B\land K)}$ \\ 
  \hline
0.2 & 0.2 & 1 & 1 & 6 & 0.33 & 0.33 \\ 
  0.2 & 0.4 & 1 & 1 & 6 & 0.33 & 0.33 \\ 
  0.2 & 0.6 & 1 & 1 & 6 & 0.33 & 0.33 \\ 
  0.2 & 0.8 & 1 & 1 & 6 & 0.33 & 0.33 \\ 
  0.2 & 1 & 1 & 1 & 6 & 0.33 & 0.33 \\ 
  0.4 & 0.2 & 1 & 1 & 15 & 0.7 & 0.67 \\ 
  0.4 & 0.4 & 1 & 1 & 15 & 0.68 & 0.67 \\ 
  0.4 & 0.6 & 1 & 1 & 15 & 2.69 & 2.67 \\ 
  0.4 & 0.8 & 1 & 1 & 15 & 2.69 & 2.67 \\ 
  0.4 & 1 & 1 & 1 & 15 & 2.67 & 2.67 \\ 
  0.6 & 0.2 & 1 & 1 & 90 & 3.61 & 4 \\ 
  0.6 & 0.4 & 1 & 1 & 225 & 6.83 & 7.04 \\ 
  0.6 & 0.6 & 1 & 1 & 225 & 16.21 & 14.08 \\ 
  0.6 & 0.8 & 1 & 1 & 90 & 21.23 & 17.97 \\ 
  0.6 & 1 & 0.67 & 1 & 15 & 24.9 & 21.97 \\ 
  0.8 & 0.2 & 1 & 1 & 270 & 7.02 & 8.98 \\ 
  0.8 & 0.4 & 0.89 & 0.95 & 1260 & 23.75 & 22.57 \\ 
  0.8 & 0.6 & 0.43 & 0.32 & 1260 & 62.98 & 38.86 \\ 
  1 & 0.2 & 0.73 & 1 & 455 & 10.36 & 20.47 \\ 
\hline
\end{tabular}
}
\end{minipage}%
\hfill
\begin{minipage}[t]{0.48\linewidth}
\centering
\caption{$Q_B$ vs min K in 9 factor fitted models}
\label{tab:right_table}
\resizebox{\linewidth}{!}{
\begin{tabular}{ccccccc} 
\hline
$\pi_1$ & $\pi_2$ & $\frac{n(Q_B)}{n_T}$ & $\frac{n(K)}{n_T}$ & $n_T$  & $\frac{\sum_{Q_B}A_s}{n(Q_B\land K)}$&$\frac{\sum_{K}A_s}{n(Q_B\land K)}$ \\ 
  \hline
0.1 & 0.1 & 1 & 1 & 9 & 0.25 & 0.25 \\ 
  0.1 & 0.3 & 1 & 1 & 9 & 0.25 & 0.25 \\ 
  0.1 & 0.5 & 1 & 1 & 9 & 0.25 & 0.25 \\ 
  0.1 & 0.7 & 1 & 1 & 9 & 0.25 & 0.25 \\ 
  0.1 & 0.9 & 1 & 1 & 9 & 0.25 & 0.25 \\ 
  0.3 & 0.1 & 1 & 1 & 84 & 0.75 & 0.75 \\ 
  0.3 & 0.3 & 1 & 0.95 & 252 & 2.37 & 2.25 \\ 
  0.3 & 0.5 & 1 & 0.95 & 252 & 3.97 & 3.75 \\ 
  0.3 & 0.7 & 1 & 0.95 & 252 & 3.98 & 3.75 \\ 
  0.3 & 0.9 & 0.98 & 0.95 & 84 & 5.59 & 5.25 \\ 
  0.5 & 0.1 & 1 & 0.9 & 756 & 2.69 & 2.5 \\ 
  0.5 & 0.3 & 0.98 & 0.83 & 1890 & 4.59 & 4 \\ 
  0.5 & 0.5 & 0.9 & 0.75 & 2520 & 6.68 & 5.5 \\ 
  0.5 & 0.7 & 0.87 & 0.7 & 1890 & 8.61 & 7 \\ 
  0.5 & 0.9 & 0.9 & 0.7 & 756 & 11.04 & 8.5 \\ 
  0.7 & 0.1 & 0.95 & 0.63 & 8820 & 5.91 & 4.5 \\ 
  0.7 & 0.3 & 0.8 & 0.29 & 114660 & 13.35 & 7.5 \\ 
  0.9 & 0.1 & 0.82 & 0.26 & 29484 & 14.21 & 6.5 \\ 
\hline
\end{tabular}
}
\end{minipage}
\end{table}

\subsection{Comparisons with restricted optimal designs}
Various literature has discussed assessing designs under certain restrictions. As an example, \cite{doi:10.1080/0740817X.2016.1241458} discussed finding new types of frequency tables to assess the quality of level balanced factorial designs.  

In the current study we have used an additional algorithm to find $Q_B$ optimal designs in the level balanced case. We start from a level balanced design and improve the $Q_B$ value by swapping the entries within the columns in a special way. In each column we divide the entries into two sets consisting of all ‘+1’ and ‘-1’ respectively. For each ‘+1’, say, we consider the $Q_B$ performance when it switches with each possible ‘-1’. If one ‘+1’ has swapped with the other ‘-1’ and the maximum decrease in $Q_B$ has been obtained we will cease to change the relative position of these two elements but to go for another ‘+1’. This continues for the rest of the columns in the starting design and we will redo the process until the design cannot be improved and then go for another starting design. The $Q_B$ value found by restricting to level balanced designs for 6 factors is given in Table \ref{tab:612balance}. By comparing this result with Table \ref{tab:re}, we find that there are 3 entries where designs imposing level-balance are inferior in terms of the $Q_B$ value for the cases in Table~\ref{tab:left_table} where the submodels considered are fully estimable.

\begin{table}[htbp]  
\caption{$Q_B$ values obtained under the level balance condition}
\label{tab:612balance}
\begin{center}
\resizebox{0.6\linewidth}{!}{
\begin{tabular}{cccccc} 

\hline
$\pi_1$ & $\pi_2$ & $Q_B$ & $\pi_1$ & $\pi_2$ & $Q_B$ \\
\hline
0.2 & 0.2 & 0.0785 & 0.8 & 0.2 & 4.1834 \\
0.2 & 0.4 & 0.1633 & 0.8 & 0.4 & 12.5533 \\
0.2 & 0.6 & 0.2586 & 0.8 & 0.6 & 21.8990 \\
0.2 & 0.8 & 0.3601 & 0.8 & 0.8 & 32.7769 \\
0.2 & 1 & 0.4693 & 0.8 & 1 & 43.4859 \\
0.4 & 0.2 & 0.5584 & 1 & 0.2 & 8.6933 \\
0.4 & 0.4 & 1.3187 & 1 & 0.4 & 23.1644 \\
0.4 & 0.6 & 2.2827 & 1 & 0.6 & 42.6311 \\
0.4 & 0.8 & 3.3649 & 1 & 0.8 & 64.9867 \\
0.4 & 1 & 4.5227 & 1 & 1 & 86.6667 \\
0.6 & 0.2 & 1.7288 &  &  & \\ 
0.6 & 0.4 & 4.8817 &  &  &  \\
0.6 & 0.6 & 8.5420 &  &  & \\ 
0.6 & 0.8 & 12.7828 &  &  & \\ 
0.6 & 1 & 17.7472 &  &  & \\
\hline

\end{tabular}
}
\end{center}
\end{table}

For the case of 9 factors, the results are given in Table~\ref{tab:916balance}. It identifies a level balanced design to be better under the $Q_B$ criterion under the prior $\pi_1=0.7,\pi_2=0.3$ for the cases inTable~\ref{tab:right_table} where the submodels considered are fully estimable. Furthermore, compared to the extended coordinate exchange algorithm, focusing on level balanced design could greatly reduce the number of starting designs that need to be used, thus improving the computational efficiency.

\begin{table}[htbp]  
\caption{$Q_B$ values obtained under the level balance condition}
\label{tab:916balance}
\begin{center}
\resizebox{0.6\linewidth}{!}{
\begin{tabular}{cccccc} 

\hline
$\pi_1$ & $\pi_2$ & $Q_B$ & $\pi_1$ & $\pi_2$ & $Q_B$ \\
\hline
0.1 & 0.1 & 0.0089 & 0.7 & 0.1 & 3.3773 \\
0.1 & 0.3 & 0.0297 & 0.7 & 0.3 & 19.4949 \\
0.1 & 0.5 & 0.0546 & 0.7 & 0.5 & 41.0571 \\
0.1 & 0.7 & 0.0835 & 0.7 & 0.7 & 68.3709 \\
0.1 & 0.9 & 0.1164 & 0.7 & 0.9 & 101.9080 \\
0.3 & 0.1 & 0.2676 & 0.9 & 0.1 & 7.6785 \\
0.3 & 0.3 & 1.0478 & 0.9 & 0.3 & 45.7507 \\
0.3 & 0.5 & 2.1546 & 0.9 & 0.5 & 99.0711 \\
0.3 & 0.7 & 3.5880 & 0.9 & 0.7 & 168.4602 \\
0.3 & 0.9 & 5.1876 & 0.9 & 0.9 & 254.8555 \\
0.5 & 0.1 & 1.2275 &  &  &  \\
0.5 & 0.3 & 5.9850 &  &  &  \\
0.5 & 0.5 & 12.9375 &  &  &  \\
0.5 & 0.7 & 20.9475 &  &  &  \\
0.5 & 0.9 & 30.5775 &  &  &  \\
\hline

\end{tabular}
}
\end{center}
\end{table}

On the other hand, \cite{lin} finds optimal designs among level balanced and orthogonal designs when they exist, i.e.\ $b_1(1 )=b_2(2 )=0$ and designs with the smallest $b_1(1 )$ and $b_2(2 )$ otherwise. For 6 factors and 12 runs we have tried to search for designs satisfying the level balance and orthogonality condition by exploring all possible $11\times6$ sub matrices from the Hadamard matrix of order 12. In terms of sequences of the generalized word counts, there are 2 distinct designs, one of which is given by the minimal K-aberration design under centered parameterization, and they all give the same $Q_B$ value as the minimum K-aberration design whose performance can be seen in Table ~\ref{tab:re}. 

For orthogonal designs with 9 factors and 16 runs, we  found $D_1$ and $D_2$ by our extended coordinate exchange algorithm where $D_1$ gives the same $Q_B$ value as the minimal K-aberration design and their performance can be seen in Table \ref{tab:916effi}.

An enumeration of non-isomorphic orthogonal arrays can be found using the orthogonal array package based on the paper \cite{Eric}. Based on the uniqueness pattern combination of $b_3(3)$ and $b_4(4)$ there are 17 distinct  generalized word count patterns, where 13 of these patterns would be inferior to $D_2$ for $Q_B$  under any prior pair $0 \leq \pi_1,\pi_2 \leq 1$. We have calculated the $Q_B$ value using prior pairs in Table \ref{tab:916effi} for the remaining 4 patterns of orthogonal arrays including $D_1$ and $D_2$ and no better $Q_B$ value has been found, thus giving reassurance that the $Q_B$ results we give in this paper are reliable.

\section{Conclusions and Discussion}
\label{sec:conclusion}
In this paper we have extended the $Q_B$ criterion to be applicable to the baseline parameterization. We have first established the association matrix that expresses the estimators of effects under the baseline parameterization in an equivalent form as a linear function of estimators of effects under the traditional centered parameterization. We are then able to generalize the $Q_B$ criterion to be applicable to the baseline parameterization. We have proved that at one given $\pi_1$, the $Q_B$ value for the minimal K-aberration design approaches 0 as $\pi_2$ converges to 0 from above. We have also found the optimal designs under the $Q_B$ criterion in the baseline parameterization for various priors under two setups of factor sizes. For 9 factors, the $A_s$ projection property given by the $Q_B$ optimal design is better than that of the minimal K-aberration design in $\frac{2}{3}$ of the cases considered. For 6 factors, the $Q_B$ optimal design performs no worse than the minimal K-aberration design in most cases and there are just 2 cases where $Q_B$ optimal design is inferior in both the $A_s$ projection and the average $A_s$ criterion.

We have also investigated the impact on the optimal designs if we impose some restrictions and we have seen that under the priors already explored focusing on level balanced designs could usually correctly recover the $Q_B$ optimal designs more efficiently than the extended coordinate exchange approach. Further imposing the condition of orthogonality would also give a $Q_B$ optimal design for a range of priors.

One issue regarding the coordinate exchange type of algorithm is that it cannot verify whether the design found is indeed globally optimal. This issue has been discussed in \cite{article}. It is worth considering  constructing approaches to prove mathematically that the $Q_B$ optimal designs we have found and those that would be found in further works are indeed optimal. We plan to look into this issue in our future work.


\section*{Acknowledgements}

The first author expresses sincere thanks for funding through the King's-China Scholarship Council Scholarship programme (File No. 202108060084).
\par


\bibhang=1.7pc
\bibsep=2pt
\fontsize{9}{14pt plus.8pt minus .6pt}\selectfont
\renewcommand\bibname{\large \bf References}
\expandafter\ifx\csname
natexlab\endcsname\relax\def\natexlab#1{#1}\fi
\expandafter\ifx\csname url\endcsname\relax
  \def\url#1{\texttt{#1}}\fi
\expandafter\ifx\csname urlprefix\endcsname\relax\def\urlprefix{URL}\fi

\bibliographystyle{chicago}      
\bibliography{reference}   

\vskip .65cm
\noindent
Xietao Zhou 
\vskip 2pt
\noindent
E-mail: (xietao.zhou@kcl.ac.uk)
\vskip 2pt

\noindent
Steven Gilmour
\vskip 2pt
\noindent
E-mail: (steven.gilmour@kcl.ac.uk)

\end{document}